\documentclass[onecolumn, amssymb, amsmath, superscriptaddress, eqsecnum, nofootinbib,notitlepage]{revtex4-1}

\usepackage{braket,verbatim}
\usepackage{graphicx,tikz,bbm,bm}
\usepackage[colorlinks=true, citecolor=blue, linkcolor=., anchorcolor=.,filecolor=.,menucolor=.,runcolor=.,urlcolor=.,]{hyperref}
\usepackage{amssymb,amsmath}

\usepackage[normalem]{ulem}

\newcommand{\bigtimes}{\mathbin{\tikz [x=1.4ex,y=1.4ex,line width=.2ex] \draw (0,0) -- (1,1) (0,1) -- (1,0);}}%

\newcommand{\be}{\begin{equation}}
\newcommand{\ee}{\end{equation}}

\newcommand{\Tr}{\text{Tr}}
\newcommand{\h}{\hat}
\newcommand{\ex}{\text{ex}}
\newcommand{\mbV}{\mathbb{V}}
\newcommand{\mbK}{\mathbb{K}}
\newcommand{\vg}{\vec{g}}
\newcommand{\mcH}{\mathcal{H}}
\newcommand{\vphi}{\varphi}
\newcommand{\mcO}{\mathcal{O}}
\newcommand{\mcV}{\mathcal{V}}

\newcommand{\su}{\mathfrak{su}}

\begin{document}








\title{Thermal quantum spacetime}




\author{Isha Kotecha}
\email{isha.kotecha@aei.mpg.de}
\affiliation{Max Planck Institute for Gravitational Physics (Albert Einstein Institute), Am M\"{u}hlenberg 1, 14476 Potsdam-Golm, Germany \\ }
\affiliation{Institut f\"{u}r Physik, Humboldt-Universit\"{a}t zu Berlin, Newtonstra{\ss}e 15, 12489 Berlin, Germany}


\begin{abstract}

The intersection of thermodynamics, quantum theory and gravity has revealed many profound insights, all the while posing new puzzles. In this article, we discuss an extension of equilibrium statistical mechanics and thermodynamics potentially compatible with a key feature of general relativity, background independence; and we subsequently use it in a candidate quantum gravity system, thus providing a preliminary formulation of a thermal quantum spacetime. Specifically, we emphasise on an information-theoretic characterisation of generalised Gibbs equilibrium that is shown to be particularly suited to background independent settings, and in which the status of entropy is elevated to being more fundamental than energy. We also shed light on its intimate connections with the thermal time hypothesis. Based on this we outline a framework for statistical mechanics of quantum gravity degrees of freedom of combinatorial and algebraic type, and apply it in several examples. In particular, we provide a quantum statistical basis for the origin of covariant group field theories, shown to arise as effective statistical field theories of the underlying quanta of space in a certain class of generalised Gibbs states.



\end{abstract}

\maketitle

\tableofcontents


\newpage

\section{Introduction}

Background independence is a hallmark of general relativity that has revolutionised our conception of space and time. The picture of physical reality it paints is that of an impartial dynamical interplay between matter and gravitational fields. Spacetime is no longer a passive stage on which matter performs; it is an equally active performer in itself. Coordinates are gauge, thus losing their physical status of non-relativistic settings. In particular, the notion of time is modified drastically. It is no longer an absolute, global, external parameter uniquely encoding the full dynamics. It is instead a gauge parameter associated with a Hamiltonian constraint. \ 


On the other hand, the well-established fields of quantum statistical mechanics and thermodynamics have been of immense use in the physical sciences. From early applications to heat engines and study of gases, to modern day uses in condensed matter systems and quantum optics, these powerful frameworks have greatly expanded our knowledge of physical systems. However, a complete extension of them to a background independent setting, such as for a gravitational field, remains an open issue \cite{Rovelli:1993ys,Connes:1994hv,Rovelli:2012nv}. The biggest challenge is the absence of an absolute notion of time, and thus of energy, which is essential to any statistical and thermodynamical consideration. This issue is particularly exacerbated in the context of defining statistical equilibrium, for the natural reason that the standard concepts of equilibrium and time are tightly linked. In other words, the constrained dynamics of a background independent system lacks a non-vanishing Hamiltonian in general, which makes formulating (equilibrium) statistical mechanics and thermodynamics, an especially thorny problem. This is a foundational issue, and tackling it is important and interesting in its own right. And even more so because it could provide deep insights into the nature of (quantum) gravitational systems. This paper is devoted to addressing precisely these points. \

The importance of addressing these issues is further intensified in light of the deep interplay between thermodynamics, gravity and the quantum theory, first uncovered for black holes. The laws of black hole mechanics \cite{Bardeen:1973gs} were a glimpse into a curious intermingling of thermodynamics and classical gravity, even if originally only at a formal level of analogy. The discovery of black hole entropy and radiation \cite{Bekenstein:1972tm,Bekenstein:1973ur,Hawking:1974sw} further brought quantum mechanics into the mix.
This directly led to a multitude of new conceptual insights along with many puzzling questions which continue to be investigated still after decades. The content of the discovery, namely that a black hole must be assigned \emph{physical} entropy and that it scales with the area of its horizon in Planck units, has birthed several distinct lines of thoughts, in turn leading to different (even if related) lines of investigations, like thermodynamics of gravity, analogue gravity and holography. Moreover, early attempts at understanding the physical origin of this entropy \cite{Bombelli:1986rw} made evident the relevance of quantum entanglement, thus also contributing to the current prolific interest in fascinating connections between quantum information theory and gravitational physics. \ 

This discovery further hinted at a quantum microstructure underlying a classical spacetime. This perspective is shared, to varying degrees of details, by various approaches to quantum gravity such as loop quantum gravity (and related spin foams and group field theories), string theory and AdS/CFT, simplicial gravity and causal set theory to name a few. Specifically within discrete non-perturbative approaches, spacetime is replaced by more fundamental entities that are discrete, quantum, and pre-geometric in the sense that no notion of smooth metric geometry and spacetime manifold exists yet. The collective dynamics of such quanta of geometry, governed by some theory of quantum gravity is then hypothesised to give rise to an emergent spacetime, corresponding to certain phases of the full theory. This would essentially entail identifying suitable procedures to extract a classical continuum from a quantum discretuum, and to reconstruct general relativistic gravitational dynamics coupled with matter (likely with quantum corrections). This emergence in quantum gravity is akin to that in condensed matter systems in which also coarse-grained macroscopic (thermodynamic) properties of the physical systems are extracted from the microscopic (statistical and) dynamical theories of the constituent atoms. In this sense our universe can be understood as an unusual condensed matter system, brought into the existing smooth geometric form by a phase transition of a quantum gravity system of pre-geometric `atoms' of space; 
in particular, as a condensate \cite{Oriti:2016acw}. \\


This brings our motivations full circle, and to the core of this article: to illustrate, the potential of and preliminary evidence for, a rewarding exchange between a background independent generalisation of statistical mechanics and discrete quantum gravity; and show that ideas from the former are vital to investigate statistical mechanics and thermodynamics of quantum gravity, and that its considerations in the latter could in turn provide valuable insights into the former. \\

These are the two facets of interest to us here. In section \ref{one}, we discuss a potential background independent extension of equilibrium statistical mechanics, giving a succinct yet complete discussion of past works in \ref{past}, and subsequently focussing on a new `thermodynamical' characterisation for background independent equilibrium in \ref{maxent}, which is based on a constrained maximisation of information entropy. In section \ref{rem} we detail further crucial properties of this characterisation, while placing it within a bigger context of the issue of background independent statistical equilibrium, also in comparison with the previous proposals. Section \ref{tth} is more exploratory, remarking on exciting new connections between the thermodynamical characterisation and the thermal time hypothesis, wherein information entropy and observer dependence are seen to play instrumental roles. In section \ref{td}, we discuss several aspects of a generalised thermodynamics based on the generalised equilibrium statistical mechanics derived above, including statements of the zeroth and first laws. Section \ref{two} is devoted to statistical mechanical considerations of candidate quantum gravity degrees of freedom of combinatorial and algebraic type. After clarifying the framework for many-body mechanics of such atoms of space in section \ref{smgeom}, we give an overview of examples in section \ref{examples}, thus illustrating the applicability of the generalised statistical framework in quantum gravity. The one case for which we give a slightly more detailed account is that of deriving a generic covariant group field theory as an effective statistical field theory starting from a particular class of quantum Gibbs states of the underlying microscopic system. Finally, we conclude and offer some outlook.



\section{Background independent equilibrium statistical mechanics} \label{one}

Covariant statistical mechanics \cite{Rovelli:1993ys,Connes:1994hv,Rovelli:2012nv} broadly aims at addressing the foundational issue of defining a suitable statistical framework for constrained systems. This issue, especially in the context of gravity, was brought to the fore in a seminal work \cite{Rovelli:1993ys}, and developed subsequently in \cite{Connes:1994hv,Rovelli:1993zz,Montesinos:2000zi,Rovelli:2012nv}. Valuable insights from these studies on spacetime relativistic systems \cite{Rovelli:1993ys,Connes:1994hv,Rovelli:2012nv,Chirco:2013zwa,Rovelli:2010mv,Montesinos:2000zi} have also formed the conceptual backbone of first applications to discrete quantum gravity \cite{Kotecha:2018gof,Chirco:2018fns,Chirco:2019kez}. In this section, we present extensions of equilibrium statistical mechanics to background independent\footnote{In the original works mentioned above, the framework is usually referred to as covariant or general relativistic statistical mechanics. But we will choose to call it background independent statistical mechanics as our applications to quantum gravity are evident of the fact that the main ideas and structures are general enough to be used in radically background independent systems devoid of any spacetime manifold or associated geometric structures.} systems, laying out different proposals for a generalised statistical equilibrium, but emphasising on one in particular, and based on which further aspects of a generalised thermodynamics are considered. The aim here is thus to address the fundamental problem of formulating these frameworks in settings where the conspicuous absence of time and energy is particularly tricky. \

Section \ref{geneqm} discusses background independent characterisations of equilibrium Gibbs states, of the general form $e^{-\sum_a \beta_a \mcO_a}$. In \ref{past}, we touch upon various proposals for equilibrium put forward in past studies on spacetime covariant systems \cite{Rovelli:1993ys,Montesinos:2000zi,Rovelli:2012nv,Josset:2015uja,Haggard:2013fx}. From section \ref{maxent} onwards, we focus on Jaynes' information-theoretic characterisation \cite{Jaynes:1957zza,Jaynes:1957zz} for equilibrium. This was first suggested as a viable proposal for background independent equilibrium, and illustrated with an explicit example in the context of quantum gravity, in \cite{Kotecha:2018gof}. Using the terminology of \cite{Kotecha:2018gof}, we call this a `thermodynamical' characterisation of equilibrium, to contrast with the customary Kubo-Martin-Schwinger (KMS) \cite{robinson} `dynamical' characterisation\footnote{For a more detailed discussion of the comparison between these two characterisations, we refer the reader to \cite{Kotecha:2018gof}. The main idea is that the various proposals for generalised Gibbs equilibrium can be divided into these two categories. Which characterisation one chooses to use in a given situation depends on the information/description of the system that one has at hand. For instance, if the description includes a 1-parameter flow of physical interest, then using the dynamical characterisation, i.e. satisfying the KMS condition with respect to it, will define equilibrium with respect to it. The procedures defining these two categories can thus be seen as `recipes' for constructing a Gibbs state; and which one is more suitable depends on our knowledge of the system.}. \

We devote section \ref{rem} to discussing various aspects of the thermodynamical characterisation, including highlighting many of its favourable features, also compared to the other proposals. In fact, we point out how this characterisation can comfortably accommodate the other proposals for Gibbs equilibrium. \

Further, as will be evident shortly, the thermodynamical characterisation hints at the idea that entropy is a central player, which has been a recurring theme across modern theoretical physics. In section \ref{tth} we present a tentative discussion on some of these aspects. In particular, we notice compelling new relations between the thermodynamical characterisation and the thermal time hypothesis, which further seem to hint at intriguing relations between entropy, observer dependence and thermodynamical time. We further propose to use the thermodynamical characterisation as a constructive criterion of choice for the thermal time hypothesis. \

Finally in section \ref{td} we define the basic thermodynamic quantities which can be derived immediately from a generalised equilibrium state, without requiring any additional physical and/or interpretational inputs. We clarify the issue of extracting a single common temperature for the full system from a set of several of them, and end with the zeroth and first laws of a generalised thermodynamics.

\subsection{Generalised equilibrium} \label{geneqm}

Equilibrium states are a cornerstone of statistical mechanics, which in turn is the theoretical basis for thermodynamics. They are vital in the description of macroscopic systems with a large number of microscopic constituents. In particular Gibbs states $e^{-\beta E}$, have a vast applicability across a broad range of fields such as condensed matter physics, quantum information and tensor networks, and (quantum) gravity, to name a few. They are special, being the unique class of states in finite systems satisfying the KMS condition\footnote{The algebraic KMS condition \cite{robinson} is well known to provide a comprehensive characterisation of statistical equilibrium in systems of arbitrary sizes, as long as there exists a well-defined 1-parameter dynamical group of automorphisms of the system. This latter point, of the required existence of a preferred time evolution of the system, is exactly the missing ingredient in our case, thus limiting its applicability.}. Furthermore, usual coarse-graining techniques also rely on the definition of Gibbs measures. In treatments closer to hydrodynamics, one often considers the full (non-equilibrium) system as being composed of many interacting subsystems, each considered to be at local equilibrium. While in the context of renormalisation group flow treatments, each phase at a given scale, for a given set of coupling parameters is also naturally understood to be at equilibrium, each described by (an inequivalent) Gibbs measure. \

Given this physical interest in Gibbs states, the question then is how to define them for background independent systems. The following are different proposals, all relying on different principles originating in standard non-relativistic statistical mechanics, extended to a relativistic setting. 

\subsubsection{Past proposals} \label{past}

The first proposal \cite{Rovelli:1993ys,Chirco:2013zwa} was based on the idea of statistical independence of arbitrary (small, but macroscopic) subsystems of the full system. The notion of equilibrium is taken to be characterised by the factorisation property of the state,  $ \rho_{12} = \rho_1 \rho_2 $, for any two subsystems 1 and 2; and the full system is at equilibrium if any one of its subsystems satisfies this property with all the rest. We notice that the property of statistical independence is related to an assumption of weak interactions \cite{Landau:1980mil}. \

This same dilute gas assumption is integral also to the Boltzmann method of statistical mechanics. It characterises equilibrium as the most probable distribution, that is one with maximum entropy\footnote{Even though this method relies on maximising the entropy like the thermodynamical characterisation, it is more restrictive than the latter, as will be made clear in section \ref{rem}.}. This method is used in \cite{Montesinos:2000zi} to study a gas of constrained particles\footnote{We remark that except for this one work, all other studies in spacetime covariant statistical mechanics are carried out from the Gibbs ensemble point of view.}. \

The work in \cite{Rovelli:2012nv} puts forward a physical characterisation for an equilibrium state. The suggestion is that, $\rho$ (itself a well-defined state on the physical, reduced state space) is said to be a physical Gibbs state if its modular Hamiltonian $h = -\ln \rho$, is a well-defined function on the physical state space; and, is such that there exists a (local) clock function $T({\bf x})$ on the extended state space (with its conjugate momentum $p_T({\bf x})$), such that the (pull-back) of $h$ is proportional to (the negative of) $p_T$. Importantly, when this is the case the modular flow (`thermal time', see section \ref{rem}) is a geometric (foliation) flow in spacetime, in which sense $\rho$ is said to be `physical'. Notice that the built-in strategy here is to define KMS equilibrium in a deparametrized system (thus it is an example of using the dynamical characterisation), since it basically identifies a state's modular Hamiltonian with a (local) clock Hamiltonian on the base spacetime manifold.  \

Another strategy \cite{Josset:2015uja} is based on the use of the ergodic principle and introduction of clock subsystems to define (clock) time averages. Again, this characterisation like a couple of the previous ones, relies on the validity of a postulate, even if traditionally a fundamental one. \

Finally, the proposal in \cite{Haggard:2013fx} interestingly characterises equilibrium by a vanishing information flow between interacting histories. The notion of information used is that of Shannon (entropy), 
$ I = \ln N $,
where $N$ is the number of microstates traversed in a given history during interaction. Equilibrium between two histories 1 and 2 is encoded in a vanishing information flow, $\delta I = I_2 - I_1 = 0$. This characterisation of equilibrium is evidently information-theoretic, even if relying on an assumption of interactions. Moreover it is much closer to our thermodynamical characterisation, because the condition of vanishing $\delta I$ is nothing but an optimisation of information entropy. \

These different proposals, along with the thermal time hypothesis \cite{Rovelli:1993ys,Connes:1994hv}, have led to some remarkable results, like recovering the Tolman-Ehrenfest effect \cite{Rovelli:2010mv,Chirco:2016wcs}, relativistic J\"uttner distribution \cite{Chirco:2016wcs} and Unruh effect \cite{Martinetti:2002sz}. However, they all assume the validity of one or more principles, postulates or assumptions about the system. Moreover, none (at least presently) seems to be general enough like the proposal below, so as to be implemented in a full quantum gravity setup, while also accommodating within it the rest of the proposals.


\subsubsection{Thermodynamical characterisation} \label{maxent}

This brings us to the proposal of characterising a generalised Gibbs state based on a constrained maximisation of information (Shannon or von Neumann) entropy  \cite{Kotecha:2018gof,Chirco:2018fns,Chirco:2019kez}, along the lines advocated by Jaynes \cite{Jaynes:1957zza,Jaynes:1957zz} purely from the perspective of evidential statistical inference. Jaynes' approach is fundamentally different from other more traditional ones of statistical physics. So too is the thermodynamical characterisation, compared with the others outlined above, as will be exemplified in the following. It is thus a new proposal for background independent equilibrium \cite{Kotecha:2018gof,Haggard:2018thr}, which has the potential of incorporating also the others as special cases, from the point of view of constructing a Gibbs state. \\

Consider a macroscopic system with a large number of constituent microscopic degrees of freedom. Our (partial) knowledge of its macrostate is given in terms of a finite set of averages $\{\langle \mcO_a \rangle = U_a\}$ of the observables we have access to. Jaynes suggests that a fitting probability estimate (which, once known, will allow us to infer also the other observable properties of the system) is not only one that is compatible with the given observations, but also that which is least-biased in the sense of not assuming any more information about the system than what we actually have at hand (namely $\{U_a\}$). In other words, given a limited knowledge of the system (which is always the case in practice for any macroscopic system), the least-biased probability distribution compatible with the given data should be preferred. As shown below, this turns out to be a Gibbs distribution with the general form $e^{-\sum_a \beta_a \mcO_a}$. \

Let $\Gamma$ be a finite-dimensional phase space (be it extended or reduced), and on it consider a finite set of smooth real-valued functions $\mcO_a$. Denote by $\rho$ a smooth statistical density (real-valued, positive and normalised function) on $\Gamma$, to be determined. Then, the prior on the macrostate gives a finite number of constraints,
\be \label{const} \langle \mcO_a \rangle_\rho = \int_{\Gamma} d\lambda \; \rho \, \mcO_a = U_a  \ee
where $d\lambda$ is a Liouville measure on $\Gamma$, and the integrals are taken to be well-defined. Further, $\rho$ has an associated Shannon entropy 
\be \label{en} S[\rho] = -\langle \ln \rho \rangle_\rho \;. \ee
By understanding $S$ to be a measure of uncertainty quantifying our ignorance about the details of the system, the corresponding bias is minimised (compatibly with the prior data) by maximising $S$ (under the set of constraints \eqref{const}, plus the normalisation condition for $\rho$) \cite{Jaynes:1957zza}. The method of Lagrange multipliers then gives a generalised Gibbs distribution of the form,
\be \label{rho} \rho_{\{\beta_a\}} = \frac{1}{Z_{\{\beta_a\}}} e^{-\sum_a \beta_a \mcO_a} \ee
where the partition function $Z_{\{\beta_a\}}$ encodes all thermodynamic properties in principle, and is assumed to be convergent. This can be done analogously for a quantum system \cite{Jaynes:1957zz}, giving a Gibbs density operator on a representation Hilbert space
\be \h{\rho}_{\{\beta_a\}}  = \frac{1}{Z_{\{\beta_a\}} } e^{-\sum_a \beta_a \h{\mcO}_a} \;.   \ee

A generalised Gibbs state can thus be defined, characterised fully by a finite set of observables of interest $\mcO_a$, and their conjugate generalised `inverse temperatures' $\beta_a$, which have entered formally as Lagrange multipliers. Given this class of equilibrium states, it should be evident that some thermodynamic quantities (like generalised `energies' $U_a$) can be identified immediately. Aspects of a generalised thermodynamics will be discussed in section \ref{td}. \

Finally, we note that the role of entropy is shown to be instrumental in defining (local\footnote{Local, in the sense of being observer-dependent (see section \ref{rem}).}) equilibrium states: ``...thus entropy becomes the primitive concept with which we work, more fundamental even than energy..." \cite{Jaynes:1957zza}. It is also interesting to notice that Bekenstein's arguments \cite{Bekenstein:1973ur} can be observed to be influenced by Jaynes' information-theoretic insights surrounding entropy, and these same insights have now guided us in the issue of background independent statistical equilibrium.


\subsection{Remarks}\label{rem}


\noindent 1. There are two key features of this characterisation. First is the use of \emph{evidential} (or epistemic, or Bayesian) probabilities, thus taking into account the given evidence $\{U_a\}$; and second is a preference for the least-biased (or most ``honest") distribution out of all the different ones compatible with the given evidence. It is not enough to arbitrarily choose any that is compatible with the prior data. An aware observer must also take into account their own ignorance, or lack of knowledge honestly, by maximising the information entropy. \\



\noindent 2. This notion of equilibrium is inherently observer-dependent because of its use of the macrostate thermodynamic description of the system, which in itself is observer-dependent due to having to choose a coarse-graining, that is the set of macroscopic observables. \\ 


\noindent 3. Given a generalised Gibbs state, the question arises as to which flow it is stationary with respect to. Any density distribution or operator satisfies the KMS condition (which implies stationarity) with respect to its own modular flow. In fact by the Tomita-Takesaki theorem \cite{robinson}, \emph{any} faithful algebraic state over a von Neumann algebra is KMS with respect to its own 1-parameter modular (Tomita) flow.\footnote{This is also the main ingredient of the thermal time hypothesis \cite{Rovelli:1993ys,Connes:1994hv}, which we will return to below.} Given this, then $\rho_{\{\beta_a\}}$ is clearly KMS with respect to the flow $X_\rho \sim \partial/\partial t$ (or $\h{U}_\rho (t) \sim e^{i\h{h}t}$) generated by its modular Hamiltonian $ h = \sum_a \beta_a \mcO_a $. In particular, $\rho_{\{\beta_a\}}$ is not stationary with respect to the individual flows $X_a$ generated by $\mcO_a$, unless they satisfy $[X_a,X_{a'}]=0$ for all $a,a'$ \cite{Chirco:2018fns}. In fact this last property shows that the proposal of \cite{Rovelli:1993ys,Chirco:2013zwa} based on statistical independence (that is $[X_{\rho_1}, X_{\rho_2}] = 0$) can be understood as a special case of this one, when the state is defined for a pair of observables that are defined on mutually exclusive subspaces of the state space. In this case, their respective flows will automatically commute and the state will be said to satisfy statistical independence.\\

\noindent 4. To be clear, the use of the `most probable' characterisation for equilibrium is not new in itself. It was used by Boltzmann in the late 19th century, and utilised (also within a Boltzmann interpretation of statistical mechanics) in a constrained system in \cite{Montesinos:2000zi}. Nor is the fact that equilibrium configurations maximise the system's entropy, which was well known already since the time of Gibbs\footnote{But as Jaynes points out in \cite{Jaynes:1957zza}, these properties were relegated to side remarks in the past, not really considered to be fundamental to the theory, nor to the justifications for the methods of statistical mechanics. }. The novelty here is: in the revival of Jaynes' perspective, of deriving equilibrium statistical mechanics in terms of evidential probabilities, solely as a problem of statistical inference without depending on the validity of any further conjectures, physical assumptions or interpretations; and, in the suggestion that it is general enough to apply to genuinely background independent systems, including quantum gravity. Below we list some of these more valuable features. 


\begin{itemize}

\item The procedure is versatile, being applicable to a wide array of cases (both classical and quantum), relying only on a sufficiently well-defined mathematical description in terms of a state space, along with a set of observables with dynamically constant averages $U_a$ defining a suitable macrostate of the system\footnote{In fact, in hindsight, we could already have anticipated a possible equilibrium description in terms of these constants, whose existence is assumed from the start.}.

\item Evidently, this manner of defining equilibrium statistical mechanics (and from it, thermodynamics) does not lend any fundamental status to energy, nor does it rely on selecting a single, special (energy) observable out of the full set $\{\mcO_a\}$. It can thus be crucial in settings where concepts of time and energy are dubious at the least, or not defined at all like in non-perturbative quantum gravity.

\item It has a technical advantage of not needing any (1-parameter) symmetry (sub-) groups of the system to be defined a priori, unlike the dynamical characterisation based on the standard KMS condition. 

\item It is independent of any additional physical assumptions, hypotheses or principles that are common to standard statistical physics, and in the present context, to the other proposals of generalised equilibrium recalled in section \ref{geneqm}. Some examples of these extra ingredients (not required in the thermodynamical characterisation) that we have already encountered are ergodicity, weak interactions, statistical independence, and often a combination of them.

\item It is independent of any physical interpretations attached (or not!) to the quantities and setup involved. This further amplifies its appeal for use in quantum gravity where the geometrical (and physical) meanings of the quantities involved may not necessarily be clear from the start. 

\item One of the main features (which helps accommodate the other proposals as special cases of this one) is the generality in the choice of observables $\mcO_a$ allowed naturally by this characterisation. In principle they need only be mathematically well-defined in the given description of the system (regardless of whether it is kinematic i.e. working at the extended state space level, or dynamic, i.e. working with the physical state space), satisfying convexity properties so that the resultant Gibbs state is normalisable. More standard choices include a Hamiltonian in a non-relativistic system, a clock Hamiltonian in a deparametrized system \cite{Kotecha:2018gof,Rovelli:2012nv}, and generators of kinematic symmetries like rotations, or more generally of 1-parameter subgroups of Lie group actions \cite{souriau,e18100370}. Some of the more unconventional choices include geometric observables like volume \cite{Kotecha:2018gof,kot}, (component functions of the) momentum map associated with geometric closure of classical polyhedra \cite{Chirco:2018fns,Chirco:2019kez}, half-link gluing (or face-sharing) constraints of discrete gravity \cite{Chirco:2018fns}, a projector in group field theory \cite{Oriti:2013aqa,Chirco:2018fns}, and generic gauge-invariant observables (not necessarily symmetry generators) \cite{Montesinos:2000zi}. We refer to \cite{Kotecha:2018gof} for a more detailed discussion.


\end{itemize}

In section \ref{examples} we outline some examples of using this characterisation in quantum gravity; while a detailed investigation of its consequences in particular for covariant systems on a spacetime manifold is left to future studies. 

 \subsection{Relation to thermal time hypothesis} \label{tth}
 
This section outlines a couple of new intriguing connections between the thermodynamical characterisation and the thermal time hypothesis, which we think are worthwhile to be explored further. Thermal time hypothesis \cite{Rovelli:1993ys,Connes:1994hv} states that the (geometric) modular flow of the (physical, equilibrium) statistical state that an observer happens to be in is the time that they experience. It thus argues for a thermodynamical origin of time \cite{Rovelli:2009ee}. \

But what is this state? Pragmatically, the state of a macroscopic system is that which an observer is able to observe and assigns to the system. It is not an absolute property since one can never know everything there is to know about the system. In other words the state that the observer `happens to be in' is the state that they are able to detect. This leads us to suggest that the thermodynamical characterisation can provide a suitable \emph{criterion of choice} for the thermal time hypothesis. \


What we mean by this is the following. Consider a macroscopic system that is observed to be in a particular macrostate in terms of a set of (constant) observable averages. The thermodynamical characterisation then provides the least biased choice for the underlying (equilibrium) statistical state. Given this state then, the thermal time hypothesis would imply that the (physical) time experienced by this observer is the (geometric) modular flow of the observed state. \

Jaynes \cite{Jaynes:1957zza,Jaynes:1957zz} had turned the usual logic of statistical mechanics upside-down to stress on entropy and the observed macrostate as the starting point, to define equilibrium statistical mechanics in its entirety \emph{from} it (and importantly, a further background independent generalisation, as we have shown above). While Rovelli \cite{Rovelli:1993ys}, later with Connes \cite{Connes:1994hv}, had turned the usual logic of the definition of time upside-down to stress on the choice of a statistical state as the starting point to identify a suitable time flow \emph{from} it. The suggestion here is to merge the two sets of insights to get an operational way of implementing the thermal time hypothesis.   \

It is interesting to see that the crucial property of observer-dependence of relativistic time arises as a natural consequence of our suggestion, directly because of the observer-dependence of any state defined using the thermodynamical characterisation. This way, thermodynamical time is intrinsically `perspectival' \cite{Rovelli:2015dha} or `anthropomorphic' \cite{Jaynes1992}. \

To be clear, this criterion of choice will not single out a preferred state, by the very fact that it is inherently observer-dependent. It is thus compatible with the basic philosophy of the thermal time hypothesis, namely that there is no preferred physical time.



Presently the above suggestion is rather conjectural, and certainly much work remains to be done to understand it better, and explore its potential consequences for physical systems. Here, it may be helpful to realise that the thermal time hypothesis can be sensed to be intimately related with (special and general) relativistic systems, and so might the thermodynamical characterisation when considered in this context. Thus for instance, Rindler spacetime or stationary black holes might offer suitable settings to begin investigating these aspects in more detail. \\

The second connection that we observe is much less direct, and is via information entropy. The generator of the thermal time flow \cite{Rovelli:1993ys}, $-\ln \rho$, can immediately be observed to be related to Shannon entropy \eqref{en}. Moreover, in the general algebraic (quantum) field theoretic setting, the generator is the log of the modular operator $\Delta$ of von Neumann algebra theory \cite{Connes:1994hv}. A modification of it, the relative modular operator, is known to be an algebraic measure of relative entropy \cite{Araki:1976zv}, which in fact has seen a recent revival in the context of quantum information and gravity. This is a remarkable feature in our opinion, which compels us to look for deeper insights it may have to offer, in further studies. 






\subsection{Generalised thermodynamic potentials, Zeroth and First laws} \label{td}

Traditional thermodynamics is the study of energy and entropy exchanges. But what is a suitable generalisation of it for background independent systems? This, like the question of a generalised equilibrium statistical mechanics which we have considered till now, is still open. In the following, we offer some insights gained from preceding discussions, including identifying certain thermodynamic potentials, and generalised zeroth and first laws. \

Thermodynamic potentials are vital, particularly in characterising the different phases of the system. The most important one is the partition function $Z_{\{\beta_a\}}$, or equivalently the free energy
\be \Phi(\{\beta_a\}) := - \ln Z_{\{\beta_a \}} \,.  \ee 
 It encodes complete information about the system from which other thermodynamic quantities can be derived in principle. Notice that the standard definition of a free energy $F$ comes with an additional factor of a (single, global) temperature, that is we normally have $\Phi = \beta F$. But for now, $\Phi$ is the more suitable quantity to define and not $F$ since we do not (yet) have a single common temperature for the full system. We will return to this point below. \

Next is the thermodynamic entropy (which by use of the thermodynamical characterisation has been identified with information entropy), which is straightforwardly
\be S(\{U_a\}) =  \sum_a  \beta_a U_a  -  \Phi   \ee
for generalised Gibbs states of the form \eqref{rho}. Notice again the lack of a single $\beta$ scaling the whole equation at this more general level of equilibrium.

By varying $S$ such that the variations $d U_a$ and $\langle d \mcO_a \rangle$ are independent \cite{Jaynes:1957zza}, a set of generalised heats can be defined
\be 
dS = \sum_a \beta_a (d U_a - \langle d \mcO_a \rangle) =: \sum_a \beta_a \,d Q_a   
\ee
and, from it (at least part of the\footnote{By this we mean that the term $\langle d\mcO_a \rangle$, based on the \emph{same} observables defining the generalised energies $U_a$, can be seen as reflecting some work done on the system. But naturally we do not expect or claim that this is all the work that is/can be performed on the system by external agencies. In other words, there could be other work contributions, in addition to the terms $dW_a$. A better understanding of work terms in this background independent setup, will also contribute to a better understanding of the generalised first law presented below.}) work done on the system $dW_a$ \cite{Chirco:2018fns}, can be identified 
\be
dW_a := \langle d\mcO_a \rangle = \frac{1}{\beta_a} \int_{\Gamma} d\lambda \; \frac{\delta \Phi}{\delta \mcO_a} \,d\mcO_a  \;.
\ee

From the setup of the thermodynamical characterisation presented in section \ref{maxent}, we can immediately identify $U_a$ as generalised ``energies''. Jaynes' procedure allows these quantities to \emph{democratically} play the role of generalised energies. None had to be selected as being \emph{the} energy in order to define equilibrium. This a priori democratic status of the several conserved quantities can be broken most easily by preferring one over the others. In turn if its modular flow can be associated with a physical evolution parameter (relational or not), then this observable can play the role of a dynamical Hamiltonian. \ 

Thermodynamic conjugates to these energies are several generalised inverse temperatures $\beta_a$. By construction each $\beta_a$ is the periodicity in the flow of $\mcO_a$, in addition to being the Lagrange multiplier for the $a^{\text{th}}$ constraint in \eqref{const}. Moreover these same constraints can  determine $\beta_a$, by inverting the equations
\be \frac{\partial \Phi}{\partial \beta_a} = U_a \,; \ee
or equivalently from
\be \frac{\partial S}{\partial U_a} = \beta_a \;. \ee
In general, $\{\beta_a\}$ is a multi-variable inverse temperature. In the special case when $\mcO_a$ are component functions of a dual vector, then $\vec{\beta} \equiv (\beta_a)$ is a vector-valued temperature. For example, this is the case when $\vec{\mcO}\equiv \{\mcO_a\}$ are dual Lie algebra-valued momentum maps associated with Hamiltonian actions of Lie groups, as introduced by Souriau \cite{souriau,e18100370}, and appearing in the context of classical polyhedra in \cite{Chirco:2018fns}. \

As we saw, a generalised equilibrium is characterised by several inverse temperatures, but an identification of a single common temperature for the full system is of obvious interest. This can be done as follows \cite{Chirco:2018fns,Chirco:2013zwa}. A state of the form \eqref{rho}, with modular Hamiltonian 
\be h = \sum_a \beta_a \mcO_a \ee
generates a modular flow (with respect to which it is at equilibrium), parametrized by
\be t = \sum_a \frac{t_a}{\beta_a} \ee
where $t_a$ are the flow parameters of $\mcO_a$. The strategy now is to reparametrize the same trajectory by a rescaling of $t$,
\be \tau := t/\beta \ee
for a real-valued $\beta$. It is clear that $\tau$ parametrizes the modular flow of a rescaled modular hamiltonian $\tilde{h} = \beta h$, associated with the state
\be \tilde{\rho}_\beta = \frac{1}{\tilde{Z}_\beta} e^{- \tilde{h}} = \frac{1}{\tilde{Z}_\beta} e^{-\beta h} \ee
characterised now by a single inverse temperature $\beta$. \

In fact, this state can be understood as satisfying the thermodynamical characterisation for a single constraint \be \langle h \rangle =  \text{constant}\ee instead of several of them \eqref{const}. Clearly, this rescaling is not a trivial move. It corresponds to the case of a weaker, single constraint which by nature corresponds to a different physical situation wherein there is exchange of information between the different observables (so that they can thermalise to a single $\beta$). This can happen for instance when one observable is special (say, the Hamiltonian) and the rest are functionally related to it (like the volume or number of particles). Whether such a determination of a single temperature can be brought about by a more physically meaningful technique is left to future work. Having said that, it will not change the general layout of the two cases as outlined above. \

One immediate consequence of extracting a single $\beta$ is regarding the free energy, which can now be written in the familiar form as
\be \Phi = \beta F \,. \ee
This is most directly seen from the expression for the entropy, 
\be \label{ds} \tilde{S} = -\langle \ln \tilde{\rho}_\beta\rangle_{\tilde{\rho}_\beta} = \beta \sum_a \beta_a \tilde{U}_a + \ln \tilde{Z} \;\;\;\Leftrightarrow\;\;\; \tilde{F} = \tilde{U} - \beta^{-1} \tilde{S} \ee
where $\tilde{U} = \sum_a \beta_a \tilde{U}_a$ is a total energy, and tildes mean that the quantities are associated with the state $\tilde{\rho}_\beta$. Notice that the above equation clearly identifies a single conjugate variable to entropy, the temperature $\beta^{-1}$.\

It is important to remark that in the above method to get a single $\beta$, we still didn't need to choose a special observable, say $\mcO'$, out of the given set of $\mcO_a$. If one were to do this, i.e. select $\mcO'$ as a dynamical energy (so that by extension the other $\mcO_a$ are functions of this one), then by standard arguments, the rest of the Lagrange multipliers will be proportional to $\beta'$, which in turn would then be the common inverse temperature for the full system. The point is that this latter instance is a special case of the former. \\

We end this section with zeroth and first laws of generalised thermodynamics. The crux of the zeroth law is a definition of equilibrium. Standard statement refers to a thermalisation resulting in a single temperature being shared by any two systems in thermal contact. This can be extended by the statement that at equilibrium, all inverse temperatures $\beta_a$ are equalised. This is in exact analogy with all intensive thermodynamic parameters, such as the chemical potential, being equal at equilibrium.  \

The standard first law is basically a statement about conservation of energy. In the generalised equilibrium case corresponding to a set of individual constraints \eqref{const}, the first law is satisfied $a^{\text{th}}$-energy-wise, 
\be dU_a = dQ_a + dW_a \;.\ee
The fact that the law holds $a$-energy-wise is not surprising because the separate constraints \eqref{const} for each $a$ mean that observables $\mcO_a$ do not exchange any information amongst themselves. If they did, then their Lagrange multipliers would no longer be mutually independent and we would automatically reduce to the special case of having a single $\beta$ after thermalisation. \

On the other hand, for the case with a single $\beta$, variation of the entropy \eqref{ds} gives
\be  d\tilde{S} = \beta \sum_a \beta_a (dU_a - \langle d\mcO_a \rangle) =: \beta d\tilde{Q} \ee
giving a first law with a more familiar form, in terms of total energy, total heat and total work variations
\be d\tilde{U} = d\tilde{Q} + d\tilde{W} \,. \ee
As before, in the even more special case where $\beta$ is conjugate to a single preferred energy, then this reduces to the traditional first law. We leave the verification of the second law for the generalised entropy to future work. Further, the quantities introduced above and the consequences of this setup need also to be investigated in greater detail.


\section{Equilibrium statistical mechanics in quantum gravity} \label{two}

Emergence of spacetime is the outstanding open problem in quantum gravity that is being addressed from several directions. One such is based on modelling quantum spacetime as a many-body system \cite{Oriti:2017twl}, which further complements the view of a classical spacetime as an effective macroscopic thermodynamic system. This formal suggestion allows one to treat extended regions of quantum spacetime as built out of discrete building blocks whose dynamics is dictated by non-local, combinatorial and algebraic mechanical models. Based on this mechanics, a formal statistical mechanics of the quanta of space can be studied \cite{Kotecha:2018gof,Chirco:2018fns}. Statistical mixtures of quantum gravity states are better suited to describe generic boundary configurations with a large number of quanta. This is in the sense that given a region of space with certain known macroscopic properties, a more reasonable modelling of its underlying quantum gravity description would be in in terms of a mixed state rather than a pure state, essentially because we cannot hope to know precisely all microscopic details to prefer one particular microstate. A simple example is having a  region with a fixed spatial volume and wanting to estimate the underlying quantum gravity (statistical) state \cite{Kotecha:2018gof,Montesinos:2000zi}. \ 

In addition to the issue of emergence, investigating the statistical mechanics and thermodynamics of quantum gravity systems would be expected to contribute towards untangling the puzzling knot between thermodynamics, gravity and the quantum theory. Especially so when applied to more physical settings, like cosmology \cite{kot}. \ 


In the rest of this article, we use results from the previous sections to outline a framework for equilibrium statistical mechanics for candidate quanta of geometry (along the lines presented in \cite{Kotecha:2018gof,Chirco:2018fns}, but generalising further to a richer combinatorics based on \cite{Oriti:2014yla}), and within it give an overview of some concrete examples. In particular, we show that a group field theory can be understood as an effective statistical field theory derived from a coarse-graining of a generalised Gibbs configuration of the underlying quanta. In addition to providing an explicit quantum statistical basis for group field theories, it further reinforces their status as being field theories for quanta of geometry \cite{Reisenberger:2000zc,Freidel:2005qe,Oriti:2006se,Oriti:2011jm}. As expected, we see that even though the many-body viewpoint makes certain techniques available that are almost analogous to standard treatments, there are several non-trivialities such as that of background independence, and physical (possible pre-geometric and effective geometric) interpretations of the statistical and thermodynamic quantities involved.


\subsection{Framework} \label{smgeom}

The candidate atoms of space considered here are geometric (quantum) $d$-polyhedra (with $d$ faces), or equivalently open $d$-valent nodes with its half-links dressed by the appropriate algebraic data \cite{Bianchi:2010gc}. This choice is motivated strongly by loop quantum gravity \cite{Bodendorfer:2016uat}, spin foam \cite{Perez:2012wv}, group field theory \cite{Reisenberger:2000zc,Freidel:2005qe,Oriti:2006se,Oriti:2011jm} and lattice quantum gravity \cite{Hamber:2009mt} approaches in the context of 4d models. Extended discrete space and spacetime can be built out of these fundamental atoms or `particles', via kinematical compositions (or boundary gluings) and dynamical interactions (or bulk bondings) respectively. In this sense the perspective innate to a many-body quantum spacetime is a constructive one, which is naturally also extended to the statistical mechanics based on this mechanics. \

Two types of data specify a mechanical model, combinatorial and algebraic. States and processes of a model are supported on combinatorial structures, here abstract\footnote{Thus not necessarily embedded into any continuum spatial manifold.} graphs and 2-complexes respectively; and algebraic dressings of these structures adds discrete geometric information. Thus, different choices of combinatorics and algebraic data gives different mechanical models. For instance, the simplest spin foam models (and their associated group field theories) for 4d gravity are based on: boundary combinatorics based on a 4-valent node (or a tetrahedron), bulk combinatorics based on a 4-simplex interaction vertex, and algebraic (or group representation) data of $SU(2)$ labelling the boundary 4-valent graphs and bulk 2-complexes. 

Clearly this is not the only choice, in fact far from it. The vast richness of possible combinatorics, compatible with our constructive point of view, is comprehensively illustrated in \cite{Oriti:2014yla}\footnote{In fact \cite{Oriti:2014yla} is phrased in a language closer to the group field theory approach, but the structures are general enough to apply elsewhere, like in spin foams, as evidenced in \cite{Finocchiaro:2018hks}.}. And the various choices for variables to label the discrete structures with (so that they may encode a suitable notion of discrete geometry, which notion depending exactly on the variables chosen and constraints imposed on them) have been an important subject of study, starting all the way from Regge \cite{Regge:1961px,Regge:2000wu,Dittrich:2008va,Freidel:2010aq,Rovelli:2010km,Dittrich:2008ar}. Accommodation of these various different choices is yet another appeal of the constructive many-body viewpoint and this framework. After clarifying further some of these aspects in the following, we will choose to work with simplicial combinatorics and $SU(2)$ holonomy-flux data for the subsequent examples.


\subsubsection{Atoms of quantum space and kinematics}

In the following we will make use of some of the combinatorial structures defined in \cite{Oriti:2014yla}. However we will be content with introducing them in a more intuitive manner, and not recalling the rigorous definitions as that will not be particularly valuable for the present discussion. The interested reader can refer to \cite{Oriti:2014yla} for details.\footnote{For clarity, we note that the terminology used here is slightly different from that in \cite{Oriti:2014yla}. Specifically the dictionary between here $\leftrightarrow$ there is: combinatorial atom or particle $\leftrightarrow$ boundary patch; interaction/bulk vertex $\leftrightarrow$ spin foam atom; boundary node $\leftrightarrow$ boundary multivalent vertex $\bar{v}$; link or full link $\leftrightarrow$ boundary edge connecting two multivalent vertices $\bar{v}_1,\bar{v}_2$; half-link $\leftrightarrow$ boundary edge connecting a multivalent vertex $\bar{v}$ and a bivalent vertex $\h{v}$. This minor difference is mainly due to a minor difference in the purpose for the same combinatorial structures. Here we are in a setup where the accessible states are boundary states, for which a statistical mechanics is defined; and the case of interacting dynamics is considered as defining a suitable (amplitude) functional over the the boundary state space. On the other hand, the perspective in \cite{Oriti:2014yla} is more in a spin foam constructive setting, so that modelling the 2-complexes as built out of fundamental spin foam atoms is more natural there.} \

\begin{figure}[t]
\includegraphics[width=3 in]{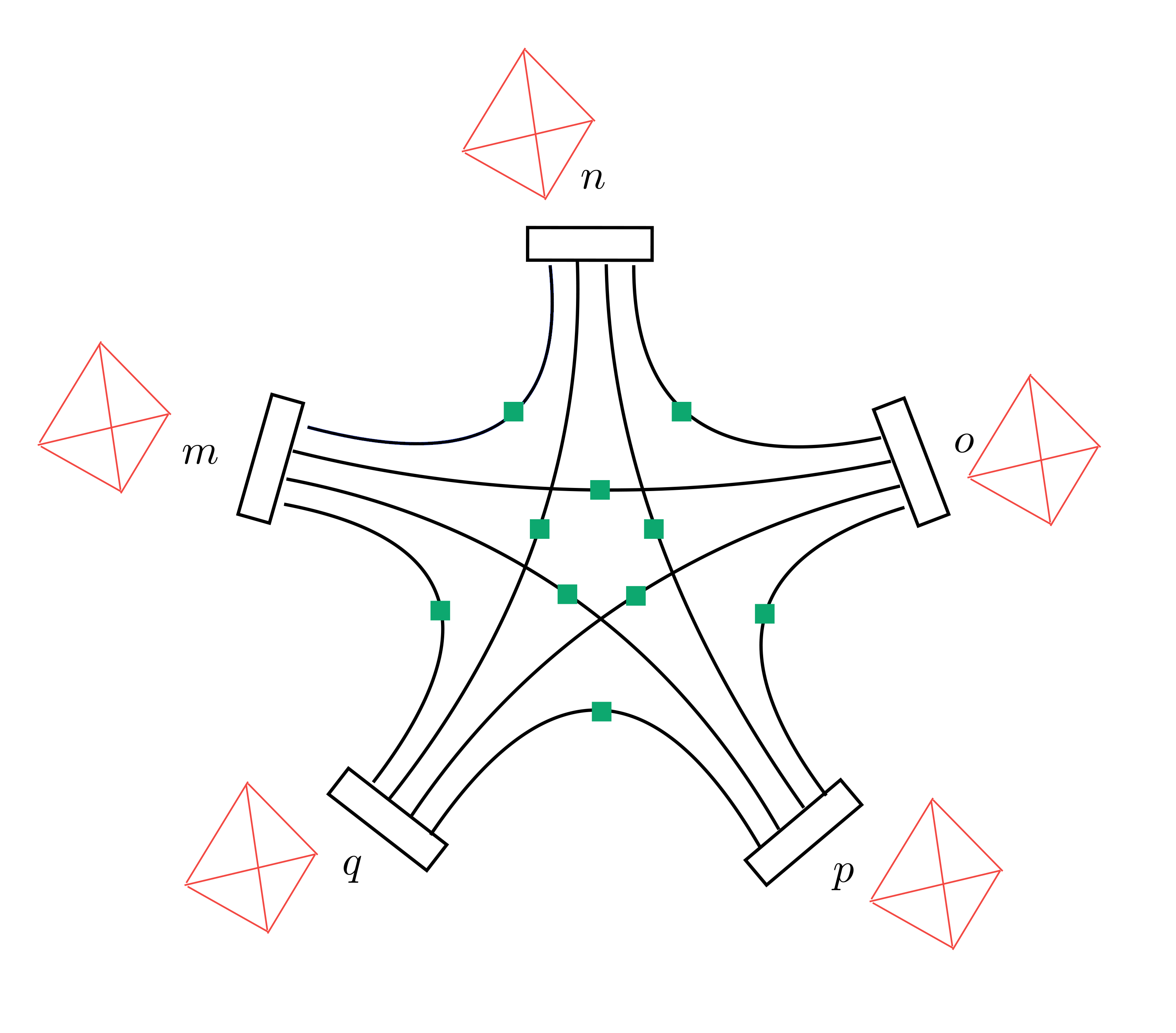}
\caption{Bisected boundary graph of a 4-simplex, as a result of non-local pair-wise gluing of half-links. Each full link is bounded by two 4-valent nodes, and bisected by one bivalent node (shown here in green).}\label{4sim}
\end{figure}

The primary objects of interest to us are boundary patches, which we take as the combinatorial atoms of space. To put simply, a boundary patch is the most basic unit of a boundary graph, in the sense that the set of all boundary patches generates the set of all connected bisected boundary graphs. A bisected boundary graph is simply a directed boundary graph with each of its full links bisected into a pair of half-links, glued at the bivalent nodes (see Figure \ref{4sim}). Different kinds of atoms of space are then the different, inequivalent boundary patches (dressed further with suitable data), and the choice of combinatorics basically boils down to a choice of the set of admissible boundary patches. Moreover, a model with multiple inequivalent boundary patches can be treated akin to a statistical system with multiple species of atoms.  \


The most general types of boundary graphs 
are those with nodes of arbitrary valence, and including loops. A common and natural restriction is to consider loopless structures, as they can be associated with combinatorial polyhedral complexes \cite{Oriti:2014yla}. As the name suggests, loopless boundary patches are those with no loops, i.e. each half-link is bounded on one end by a unique bivalent node (and on the other by the common, multivalent central node). A loopless patch is thus uniquely specified by the number of incident half-links (or equivalently, by the number of bivalent nodes bounding the central node). A $d$-patch, with $d$ number of incident half-links, is simply a $d$-valent node. Importantly for us, it is the combinatorial atom that supports (quantum) geometric states of a $d$-polyhedron \cite{Bianchi:2010gc,Barbieri:1997ks,Baez:1999tk}. A further common restriction is to consider graphs with nodes of a single, fixed valence, that is to consider $d$-regular loopless structures. \

Let's take an example. Consider the boundary graph of a 4-simplex as shown in Figure \ref{4sim}. The fundamental atom or boundary patch is a 4-valent node. This graph can be constructed starting from five open 4-valent nodes (denoted $m,n,...,q$), and gluing the half-links, or equivalently the faces of the dual tetrahedra, pair-wise, with the non-local combinatorics of a complete graph on five 4-valent nodes. The result is ten bisected full links, bounded by five nodes. It is important to note here that a key ingredient of constructing extended boundary states from the atoms are precisely the half-link gluing, or face-sharing conditions on the algebraic data decorating the patches. For instance, in the case of standard LQG holonomy-flux variables of $T^*(SU(2))$, the face-sharing gluing constraints are area matching \cite{Freidel:2010aq}, thus lending a notion of discrete classical twisted geometry to the graph. This is much weaker than a Regge geometry, which could have been obtained for the same variables if instead the so-called shape-matching conditions \cite{Dittrich:2008va} are imposed on the pair-wise gluing of faces/half-links. Thus, kinematic composition (boundary gluings) that creates boundary states depends on two crucial ingredients, the combinatorial structure of the resultant boundary graph, and face-sharing gluing conditions on the algebraic data.
 
From here on we will restrict to a single boundary patch for simplicity, a (gauge-invariant) 4-valent node dressed with $SU(2)$ data, that is a quantised tetrahedron \cite{Bianchi:2010gc,Barbieri:1997ks}. But it should be clear from the brief discussion above (and the extensive study in \cite{Oriti:2014yla}) that a direct generalisation of the present (statistical) mechanical framework is possible also for these more enhanced combinatorial structures. \

The phase space of a single classical tetrahedron, encoding both intrinsic and extrinsic degrees of freedom (along with an arbitrary orientation in $\mathbb{R}^3$) is
\be \Gamma = T^*(SU(2)^{ 4}/SU(2)) \ee
where the quotient by $SU(2)$ imposes geometric closure of the tetrahedron. The choice of domain space is basically the choice of algebraic data. For instance, in Euclidean 4d settings a more apt choice would be the group $Spin(4)$, and $SL(2,\mathbb{C})$ for Lorentzian settings. Then states of a system of $N$ tetrahedra belong to
$ \Gamma_N = \Gamma^{\times N} $,
and observables would be smooth (real-valued) functions defined on $\Gamma_N$. \cite{Kotecha:2018gof,Chirco:2018fns} \

The quantum counterparts are,
\be \mcH = L^2(SU(2)^{ 4}/SU(2)) \ee
for the single-particle Hilbert space, and 
$ \mcH_N = \mcH^{\otimes N} $
for an $N$-particle system. In the quantum setting we can go a step further and construct a Fock space based on the above single-particle Hilbert space,
\be \mcH_F = \bigoplus_{N \geq 0} \text{sym} \, \mcH_N \ee
where the symmetrisation of $N$-particle spaces implements a choice of bosonic statistics for the quanta, mirroring the graph automorphism of node exchanges. One choice for the algebra of operators on $\mcH_F$ is the von Neumann algebra of bounded linear operators. A more common choice though is the larger *-algebra generated by ladder operators $\h{\vphi},\h{\vphi}^\dag$, which generate the full $\mcH_F$ by acting on a cyclic Fock vacuum, and satisfy a commutation relations algebra 
\be [\h{\vphi}(\vec{g}_1),\h{\vphi}^\dag(\vec{g}_2)] = \int_{SU(2)} dh \prod_{I=1}^4 \delta(g_{1I}h g_{2I}^{-1})  \ee
where $\vec{g} \equiv (g_I) \in SU(2)^4$ and the integral on the right ensures $SU(2)$ gauge invariance. In fact this is the Fock representation of an algebraic bosonic group field theory defined by a Weyl algebra \cite{Kotecha:2018gof,Oriti:2013aqa,Kegeles:2017ems}. 


\subsubsection{Interacting quantum spacetime and dynamics}

Coming now to dynamics, the key ingredients here are the specifications of propagators and admissible interaction vertices, including both their combinatorics, and functional dependences on the algebraic data i.e. their amplitudes. \

The combinatorics of propagators and interaction vertices can be packaged neatly within two maps defined in \cite{Oriti:2014yla}, the bonding map and the bulk map respectively. A bonding map is defined between two bondable boundary patches. Two patches are bondable if they have the same number of nodes and links. Then, a bonding map between two bondable patches identifies each of their nodes and links, under the compatibility condition that if a bounding bivalent node in one patch is identified with a particular one in another, then their respective half-links (attaching them to their respective central nodes) are also identified with each other. So a bonding map basically bonds two bulk vertices via (parts of) their boundary graphs to form a process (with boundary). This is simply a bulk edge, or propagator. \

The set of interaction vertices can themselves be defined by a bulk map. This map augments the set of constituent elements (multivalent nodes, bivalent nodes, and half-links connecting the two) of \emph{any} bisected boundary graph, by one new vertex (the bulk vertex), a set of links joining each of the original boundary nodes to this vertex, and a set of two dimensional faces bounded by a triple of the bulk vertex, a multivalent boundary node and a bivalent boundary node. The resulting structure is an interaction vertex with the given boundary graph.\footnote{An interesting aspect is that the bulk map is one-one, so that for every distinct bisected boundary graph, there is a unique interaction vertex which can be defined from it.} The complete dynamics is then given by the chosen combinatorics, supplemented with amplitude functions that reflect the dependence on the algebraic data. \

The interaction vertices can in fact be described by vertex operators on the Fock space in terms of the ladder operators. An example vertex operator, corresponding to the 4-simplex boundary graph shown in Figure \ref{4sim}, is
\be \h{\mbV}_{\text{4sim}} = \int_{SU(2)^{20}} [dg]\;  \h{\vphi}^\dag(\vg_1) \h{\vphi}^\dag(\vg_2) V_{\text{4sim}}(\vg_1,...,\vg_5) \h{\vphi}(\vg_3) \h{\vphi}(\vg_4) \h{\vphi}(\vg_5)  \ee
where the interaction kernel $V_{\text{4sim}} = V_{\text{4sim}}(\{g_{ij}g_{ji}^{-1}\}_{i<j})$ (for $i,j=1,...,5$) encodes the combinatorics of the boundary graph. There are of course other vertex operators associated with the \emph{same} graph (that is with the same kernel), but including different combinations of creation and annihilation operators\footnote{This would generically be true for any second quantised operator \cite{Oriti:2013aqa}.}. \\



So, a definition of kinematics entails: defining the state space, which includes specifying the combinatorics (choosing the set of allowed boundary patches, which generate the admissible boundary graphs), and the algebraic data (choosing variables to characterise the discrete geometric states supported on the boundary graphs); and, defining the algebra of observables acting on the state space. A definition of dynamics entails: specifying the propagator and bulk vertex combinatorics and amplitudes. Together they specify the many-body mechanics. 


\subsubsection{Generalised equilibrium states}

Outlined below is a generalised equilibrium statistical mechanics for these systems \cite{Kotecha:2018gof,Chirco:2018fns}, along the lines laid out in section \ref{one}. For a system of many classical tetrahedra (in general, polyhedra), a statistical state $\rho_N$ can be formally defined on the state space $\Gamma_N$. If it satisfies the thermodynamical characterisation with respect to a set of functions on $\Gamma_N$ then it will be an equilibrium state. Further, a configuration with a varying number of tetrahedra can be described by a grand-canonical type state \cite{Chirco:2018fns} of the form
\be Z = \sum_{N \geq 0} e^{\mu N} Z_N \ee
where $Z_N = \int_{\Gamma_N} d\lambda \, \rho_N$, and $\mu$ is a chemical potential. Similarly for a system of many quantum tetrahedra, a generic statistical state $\h{\rho}$ is a density operator on $\mcH_F$; and generalised equilibrium states with a varying number of quanta are 
\be Z = \Tr_{\mcH_F}(e^{-\sum_a \beta_a \h{\mcO}_a + \mu \h{N}})\ee
where $\h{N} = \int d\vg \, \h{\vphi}^\dag(\vg) \h{\vphi}(\vg)$ is the number operator on $\mcH_F$. Operators of natural interest here are the ones encoding the dynamics, that is vertex (and kinetic) operators (see section \ref{examples} below). Such grand-canonical type boundary states are important because one would expect quantum gravity dynamics to not be number conserving in general \cite{Chirco:2018fns,Oriti:2013aqa}. Also, naturally in both cases, what the precise content of equilibrium is depends crucially on which observables $\mcO_a$ are used to define the state. And as pointed out in section \ref{rem}, and exemplified in the cases below in \ref{examples}, there are many choices and types of observables one could consider in principle. Which ones are the relevant ones in a given situation is in fact a crucial part of the problem.


\subsection{Applications}\label{examples}

We briefly sketch below some examples of applying the above framework. \\ 

A couple of examples for a classical system are studied in \cite{Chirco:2018fns}. In the process of applying the thermodynamical characterisation, these cases introduce a statistical, effective manner of imposing a given (set of) first class constraint(s), that is $\langle C \rangle = 0$, instead of the exact, strong way $C=0$. In one case, the condition of closure of a classical $d$-polyhedron is relaxed in this statistical manner, while in the other the boundary gluing constraints amongst the polyhedral atoms of space are relaxed in this way to describe fluctuating twisted geometries. Brief summaries of these follow. \

In the first example, starting from the extended state space $\Gamma_\ex = \bigtimes_I \, S^2_{A_I}$ of intrinsic geometries of a $d$-polyhedron with face areas $\{A_I\}_{I=1,...,d}$, closure is implemented via the following $\su(2)^*$-valued function on $\Gamma_\ex$,
\be J = \sum_{I=1}^d x_I \ee
which is the momentum map associated with the diagonal action of $SU(2)$. Satisfying closure exactly is to have $J = 0$. Then applying the thermodynamic characterisation to the scalar component functions of $J$, that is requiring $\langle J_a \rangle = 0$ $(a=1,2,3)$, gives a Gibbs distribution on $\Gamma_\ex$ of the form $e^{-\sum_a \beta_a J_a}$ with a vector-valued temperature $(\beta_a) \in \su(2)$. Thus we have a thermal state for a classical polyhedron that is fluctuating in terms of its closure, with the fluctuations controlled by the parameter $\beta$. In fact this state generalises Souriau's Gibbs states \cite{souriau,e18100370} to the case of Lie group (Hamiltonian) actions associated with first class constraints. 

In the other example, the set of half-link gluing (or face-sharing) conditions for a boundary graph are statistically relaxed. It is known that an oriented (closed) boundary graph $\gamma$, with $M$ nodes and $L$ links, labelled with $(g,x)\in T^*(SU(2))$ variables admits a notion of discrete (closed) twisted geometry \cite{Freidel:2010aq}. Twisted geometries are a generalisation of the more rigid Regge geometries, wherein the shapes of the shared faces are left arbitrary and only their areas are constrained to match. From the present constructive many-body viewpoint, one can understand these states instead as a result of satisfying a set of $SU(2)$- and $\su(2)^*$-valued gluing conditions (denoted respectively by $\{C\}$ and $\{D\}$) on an initially disconnected system of several labelled open nodes. That is, starting from a system of $M$ number of labelled open nodes, one ends up with a twisted geometric configuration if the set of gluing constraints on the holonomy and flux variables corresponding to a given $\gamma$, $\{C_{\ell,a}(g_{n\ell}g^{-1}_{m\ell}) = 0,D_{\ell,a}(x_{n\ell}-x_{m\ell}) = 0\}_\gamma$, are satisfied strongly (component-wise). Here $\ell = 1,2,...,L$ labels a full link, $a=1,2,3$ is $SU(2)$ component index, and subscripts $n\ell$ refer to the half-link (belonging to the full link) $\ell$ of node $n$. We can then choose instead to impose these constraints weakly by requiring only its statistical averages in a state to vanish. This gives a $\gamma$-dependent state on $\Gamma_{M}$, written formally as
\be \rho_{\{\gamma,\alpha,\beta\}} \propto e^{-\sum_{\ell}\sum_{a}\alpha_{\ell,a}C_{\ell,a} + \beta_{\ell,a}D_{\ell,a}} \equiv e^{-G_\gamma(\alpha,\beta)}  \ee 
where $\alpha,\beta \in \mathbb{R}^{3L}$ are generalised inverse temperatures characterising this fluctuating twisted geometric configuration. In fact, one can generalise this state to a probabilistic superposition of such internally fluctuating twisted geometries for an $N$ particle system (thus defined on $\Gamma_N$), which includes contributions from distinct graphs, each composed of a possibly variable number of nodes $M$. A state of this kind can formally be written as, 
\be \label{trho} \rho_N = \frac{1}{Z_N(M_{\max},\lambda_\gamma,\alpha,\beta)} e^{-\sum_{M=2}^{M_{\max}} \sum_{\{\gamma\}_M} \frac{1}{\text{Aut}(\gamma)}\lambda_\gamma \sum_{i_1 \neq ... \neq i_M =1}^{N} G_\gamma (\vec{g}_{i_1},\vec{x}_{i_1},...,\vec{g}_{i_M},\vec{x}_{i_M}; \alpha,\beta)} \ee
where $i$ is the particle index, and $M_{\max} \leq N$. The value of $M_{\max}$ and the set $\{\gamma\}_M$ for a fixed $M$ are model-building choices. The first sum over $M$ includes contributions from all admissible (depending on the model, determined by $M_{\max}$) different $M$-particle subgroups of the full $N$ particle system, with the gluing combinatorics of various different boundary graphs with $M$ nodes. The second sum is a sum over all admissible boundary graphs $\gamma$, with a given fixed number of nodes $M$. And, the third sum takes into account all $M$-particle subgroup gluings (according to a given fixed $\gamma$) of the full $N$ particle system. We note that the state \eqref{trho} is a further generalisation of that presented in \cite{Chirco:2018fns}, specifically the latter is a special case of the former for the case of a single term $M=M_{\max}=N$ in the first sum. Further allowing for the system size to vary, that is considering a variable $N$ gives the most general configuration, with a set of coupling parameters linked directly to the underlying microscopic model,
\be Z(M_{\max},\lambda_\gamma,\alpha,\beta) = \sum_{N \geq 0} e^{\mu N} Z_N(M_{\max},\lambda_\gamma,\alpha,\beta)\,. \ee \

A physically more interesting example is considered in \cite{Kotecha:2018gof}, which defines a thermal state with respect to a spatial volume operator,
\be \h{\rho} = \frac{1}{Z} e^{-\beta \h{\mcV}} \ee
where $\h{\mcV} = \int d\vg \; v(\vg) \h{\vphi}^\dag(\vg)\h{\vphi}(\vg)$ is a positive, self-adjoint operator on $\mcH_F$, and the state is a well-defined density operator on the same. In fact with a grand-canonical extension of it, this system can be shown to naturally support Bose-Einstein condensation to a low-spin phase \cite{Kotecha:2018gof}. Clearly, this state encodes thermal fluctuations in the volume observable, which is especially an important one in the context of cosmology. In fact the rapidly developing field of condensate cosmology \cite{universe5060147} for atoms of space of the kind considered here, is based on modelling the underlying system as a condensate, and subsequently extracting effective physics from it. These are certainly crucial steps in the direction of obtaining cosmological physics from quantum gravity \cite{Oriti:2016acw}. It is equally crucial to enrich further the microscopic quantum gravity description itself, and extract effective physics for these new cases. One such important case is to consider thermal fluctuations of the gravitational field at early times, during which our universe is expected to be in a quantum gravity regime. That is, to consider \emph{thermal} quantum gravity condensates using the frameworks laid out in this article (as opposed to the zero temperature condensates that have been used till now), and subsequently derive effective physics from them. This case would then directly reflect thermal fluctuations of gravity as being of a proper quantum gravity origin. This is investigated in \cite{kot}. \\






We end this section by making a direct link to the definition of group field theories using the above framework. Group field theories (GFT) \cite{Freidel:2005qe,Oriti:2006se,Oriti:2011jm} are non-local field theories defined over (copies of) a Lie group. Most widely studied (Euclidean) models are for real or complex scalar fields, over copies of $SU(2), Spin(4)$ or $SO(4)$. For instance, a complex scalar GFT over $SU(2)$ is defined by a partition function of the following general form,  
\be \label{zgft} Z_{\text{GFT}} = \int [D\mu(\vphi,\bar{\vphi})] \;e^{-S_{\text{GFT}}[\vphi,\bar{\vphi}]} \ee
where $\mu$ is a functional measure which in general is ill-defined, and $S_{\text{GFT}}$ is the GFT action of the form (for commonly encountered models), \be S_{\text{GFT}} = \int_{G} dg_1 \int_G dg_2 \, K(g_1,g_2) \bar{\vphi}(g_1) \vphi (g_2) + \int_G dg_1 \int_G dg_2 \,...\; V(g_1,g_2,...) f(\varphi,\bar{\varphi}) \ee
where $g\in G$, and the kernel $V$ is generically non-local, which convolutes the arguments of several $\vphi$ and $\bar{\vphi}$ fields (written here in terms of a single function $f$). It defines the interaction vertex of the dynamics by enforcing the combinatorics of its corresponding (unique, via the inverse of the bulk map) boundary graph. \

$Z_{\text{GFT}}$ defines the covariant dynamics of the GFT model encoded in $S_{\text{GFT}}$. Below we outline a way to derive such covariant dynamics from a suitable quantum statistical equilibrium description of a system of quanta of space defined previously in \ref{smgeom}. The following technique of using field coherent states is the same as in \cite{Oriti:2013aqa,Chirco:2018fns}, but with the crucial difference that here we do not claim to define, or aim to achieve any correspondence (even if formal) between a canonical dynamics (in terms of a projector operator) and a covariant dynamics (in terms of a functional integral). 
Here we simply show a quantum statistical basis for the covariant dynamics of a GFT, and in the process, reinterpret the standard form of the GFT partition function \eqref{zgft} as that of an effective statistical field theory arising from a coarse-graining and further approximations of the underlying statistical quantum gravity system. \
  
We saw in \ref{smgeom} that the dynamics of the polyhedral atoms of space is encoded in the choices of propagators and interaction vertices, which can be written in terms of kinetic and vertex operators in the Fock description. In our present considerations with a single type of atom ($SU(2)$-labelled 4-valent node), let us then consider the following generic kinetic and vertex operators,
\be \h{\mbK} = \int_{SU(2)^8} [dg] \; \h{\vphi}^\dag (\vg_1) K(\vg_1,\vg_2) \h{\vphi}(\vg_2) \;\;\;\;,\;\;\;\;
 \h{\mbV} = \int_{SU(2)^{4N}} [dg] \; V_{\gamma}(\vg_1,...,\vg_N) \h{f}(\h{\vphi},\h{\vphi}^\dag) \ee
where $N > 2$ is the number of 4-valent nodes in the boundary graph $\gamma$, and $\h{f}$ is a function of the ladder operators with all terms of a single degree $N$. For example when $N=3$, this function could be $\h{f} = \lambda_1 \h{\vphi}\h{\vphi}\h{\vphi}^\dag + \lambda_2 \h{\vphi}^\dag \h{\vphi}\h{\vphi}^\dag$. As we saw before, in principle a generic model can include several distinct vertex operators. Even though what we have considered here is the simple of case of having only one, the argument can be extended directly to the general case. \

Operators $\h{\mbK}$ and $\h{\mbV}$ have well-defined actions on the Fock space $\mcH_F$. Using the thermodynamical characterisation then, we can consider the formal constraints\footnote{A proper interpretation of these constraints is left for future work.} $\langle \h{\mbK} \rangle = $ constant and $\langle \h{\mbV} \rangle =$ constant, to write down a generalised Gibbs state on $\mcH_F$,
\be \h{\rho}_{\{\beta_a\}} = \frac{1}{Z_{\{\beta_a\}}} e^{-\beta_1 \h{\mbK} - \beta_2\h{\mbV}} \ee
where $a=1,2$ and the partition function\footnote{This partition function will in general be ill-defined as expected. One reason is the operator norm unboundedness of the ladder operators.} is, 
\be \label{zsm} Z_{\{\beta_a\}} = \Tr_{\mcH_F} (e^{-\beta_1 \h{\mbK} - \beta_2\h{\mbV}}) \;. \ee

An effective field theory can then be extracted from the above by using a basis of coherent states on $\mcH_F$ \cite{Oriti:2013aqa,Chirco:2018fns,klauderbook}. Field coherent states give a continuous representation on $\mcH_F$ where the parameter labelling each state is a wave (test) function \cite{klauderbook}. For the Fock description mentioned in section \ref{smgeom}, the coherent states are
\be \ket{\psi} = e^{\h{\varphi}^\dag(\psi) - \h{\varphi}(\psi)}\ket{0} \ee
where $\ket{0}$ is the Fock vacuum (satisfying $\h{\vphi}(\vg) \ket{0} = 0$ for all $\vg$), $\h{\vphi}(\psi) = \int_{SU(2)^4} \bar{\psi}\h{\vphi}$ and its adjoint are smeared operators, and $\psi \in \mcH$. The set of all such states provides an over-complete basis for $\mcH_F$. The most useful property of these states is that they are eigenstates of the annihilation operator, \be \h{\vphi}(\vg)\ket{\psi} = \psi(\vg)\ket{\psi} \,. \ee 
The trace in the partition function \eqref{zsm} can then be evaluated in this basis,
\begin{align} Z_{\{\beta_a\}} = \int [D\mu(\psi,\bar{\psi})] \, \bra{\psi} e^{-\beta_1 \h{\mbK} - \beta_2\h{\mbV}}\ket{\psi}  \end{align}
where $\mu$ here is the coherent state measure \cite{klauderbook}. The integrand can be treated and simplified along the lines presented in \cite{Chirco:2018fns} (to which we refer for details), to get an effective partition function, 
\be Z_0 = \int [D\mu(\psi,\bar{\psi})] \; e^{-\beta_1K[\bar{\psi},\psi] - \beta_2 V[\bar{\psi},\psi]} = Z_{\{\beta_a\}} - Z_{\mcO(\hbar)} \ee
where subscript 0 indicates that we have neglected higher order terms, collected inside $Z_{\mcO(\hbar)}$, resulting from normal orderings of the exponent in $Z_{\{\beta_a\}}$, and the functions in the exponent are $K = \bra{\psi}:\h{\mbK}:\ket{\psi}$ and $V = \bra{\psi}:\h{\mbV}:\ket{\psi}$. It is then evident that $Z_0$ has the precise form of a generic GFT partition function. It thus \emph{defines} a group field theory as an effective statistical field theory, that is \be Z_{\text{GFT}} := Z_0 \,. \ee   

From this perspective, it is clear that the generalised inverse temperatures (which are basically the intensive parameters conjugate to the energies in the generalised thermodynamics setting of \ref{td}) \emph{are} the coupling parameters defining the effective model, thus characterising the phases of the emergent statistical group field theory, as would be expected. Moreover, from this purely statistical standpoint, we can understand the GFT action more appropriately as Landau-Ginzburg free energy (or effective `Hamiltonian', in the sense that it encodes the effective dynamics), instead of a Euclidean action which might imply having Wick rotated a Lorentzian measure, even in an absence of any such notions as is the case presently. Lastly, deriving like this the covariant definition of a group field theory, based entirely on the framework presented in \ref{smgeom}, strengthens the statement that a group field theory is a field theory of combinatorial and algebraic quanta of space \cite{Oriti:2006se,Oriti:2011jm}.


\section{Conclusion and Outlook}

We have presented an extension of equilibrium statistical mechanics for background independent systems, based on a collection of results and insights from old and new studies. While various proposals for a background independent notion of statistical equilibrium have been summarised, one in particular, based on the constrained maximisation of information entropy has been stressed upon. We have argued in favour of its potential by highlighting its many unique and valuable features. We have remarked on interesting new connections with the thermal time hypothesis, in particular suggesting to use this particular characterisation of equilibrium as a criterion of choice for the application of the hypothesis. Subsequently, aspects of a generalised framework for thermodynamics have been investigated, including defining the essential thermodynamic potentials, and discussing generalised zeroth and first laws. \

We have then considered the statistical mechanics of a candidate quantum gravity system, composed of many atoms of space. The choice of (possibly different types of) these quanta is inspired directly from boundary structures in loop quantum gravity, spin foam and group field theory approaches. They are combinatorial building blocks (or boundary patches) of graphs, labelled with suitable algebraic data encoding discrete geometric information, with their constrained many-body dynamics dictated by bulk bondings between interaction vertices and amplitude functions. Generic statistical states can then be defined on a many-body state space, and generalised Gibbs states can be defined using the thermodynamical characterisation \cite{Kotecha:2018gof}. Finally, we have given an overview of applications in quantum gravity \cite{Kotecha:2018gof,Chirco:2018fns,Chirco:2019kez,kot}. In particular, we have derived the covariant definition of group field theories as a coarse-graining using coherent states of a class of generalised Gibbs states of the underlying system with respect to dynamics-encoding kinetic and vertex operators; and in this way reinterpreted the GFT partition function as an effective statistical field theory partition function, extracted from an underlying statistical quantum gravity system. \\

More investigations along these directions will certainly be worthwhile. For example, the thermodynamical characterisation could be applied in a spacetime setting, like for stationary black holes with respect to the mass, charge and angular momentum observables, to explore further its physical implications. The black hole setting could also help unfold how the selection of a single preferred temperature can occur starting from a generalised Gibbs measure. Moreover, it could offer insights into relations with the thermal time hypothesis, and help better understand some of our more intuitive reasonings presented in \ref{tth}. Similarly for generalised thermodynamics. It requires further development, particularly for the first and second laws. For instance in the first law as presented above, the additional possible work contributions need to be identified and understood, particularly in the context of background independence. For these, and other thermodynamical aspects, we could benefit from Souriau's generalisation of Lie group thermodynamics \cite{souriau,e18100370}. \

There are many avenues to explore also in the context of statistical mechanics and thermodynamics of quantum gravity. In the former, for example, it would be interesting to study potential black hole quantum gravity states \cite{Oriti:2018qty}. In general, it is important to be able to identify suitable observables to characterise an equilibrium state of physically relevant cases. On the cosmological side for instance, those phases of the complete quantum gravity system which admit a cosmological interpretation will be expected to have certain symmetries whose associated generators could then be suitable candidates for the generalised energies. Another interesting cosmological aspect to consider is that of inhomogeneities induced by early time volume thermal fluctuations of quantum gravity origin, possibly from an application of the volume Gibbs state \cite{Kotecha:2018gof} (or a suitable modification of it) recalled above. The latter aspect of investigating thermodynamics of quantum gravity would certainly benefit from confrontation with studies on thermodynamics of spacetime in semiclassical settings. We may also need to consider explicitly the quantum nature of the degrees of freedom, and use insights from the field of quantum thermodynamics \cite{doi:10.1080/00107514.2016.1201896}, which itself has fascinating links to quantum information \cite{Goold_2016}.

\begin{acknowledgments}
Many thanks are due to Daniele Oriti for valuable discussions and comments on the manuscript. Special thanks are due also to Goffredo Chirco and Mehdi Assanioussi for insightful discussions. The generous hospitality of the Visiting Graduate Fellowship program at Perimeter Institute, where part of this work was carried out, is gratefully acknowledged. Research at Perimeter Institute is supported in part by the Government of Canada through the Department of Innovation, Science and Economic Development Canada and by the Province of Ontario through the Ministry of Economic Development, Job Creation and Trade.
\end{acknowledgments}


\bibliographystyle{unsrt}
\bibliography{refSpIss}

\end{document}